\documentstyle[epsfig]{mn2e}

\begin{document}

\title[Halo-model signatures from SDSS LRGs]{Halo-model signatures
  from 380{,}000 SDSS Luminous Red Galaxies with photometric
  redshifts}

\author[Chris Blake et al.]{Chris Blake$^{1,2,}$\footnotemark, Adrian
  Collister$^3$, Ofer Lahav$^4$ \\ \\ $^1$ Centre for Astrophysics \&
  Supercomputing, Swinburne University of Technology, P.O. Box 218,
  Hawthorn, VIC 3122, Australia \\ $^2$ Department of Physics \&
  Astronomy, University of British Columbia, 6224 Agricultural Road,
  Vancouver, B.C., V6T 1Z1, Canada \\ $^3$ Institute of Astronomy,
  University of Cambridge, Cambridge, CB3 0HA, U.K. \\ $^4$ Department
  of Physics \& Astronomy, University College London, Gower Street,
  London, WC1E 6BT, U.K.}

\maketitle

\begin{abstract}
  We analyze the small-scale clustering in ``MegaZ-LRG'', a large
  photometric-redshift catalogue of Luminous Red Galaxies extracted
  from the imaging dataset of the Sloan Digital Sky Survey.
  MegaZ-LRG, presented in a companion paper, spans the redshift range
  $0.4 < z < 0.7$ with an r.m.s.\ redshift error $\sigma_z \approx
  0.03(1+z)$, covering $5{,}914$ deg$^2$ to map out a total cosmic
  volume $2.5$ $h^{-3}$ Gpc$^3$.  In this study we use $380{,}000$
  photometric redshifts to measure significant deviations from the
  canonical power-law fit to the angular correlation function in a
  series of narrow redshift slices, in which we construct
  volume-limited samples.  These deviations are direct signatures of
  the manner in which these galaxies populate the underlying network
  of dark matter haloes.  We cleanly delineate the separate
  contributions of the ``1-halo'' and ``2-halo'' clustering terms and
  fit our measurements by parameterizing the halo occupation
  distribution $N(M)$ of the galaxies.  Our results are successfully
  fit by a ``central'' galaxy contribution with a ``soft'' transition
  from zero to one galaxies, combined with a power-law ``satellite''
  galaxy component, the slope of which is a strong function of galaxy
  luminosity.  The large majority of galaxies are classified as
  central objects of their host dark matter haloes rather than
  satellites in more massive systems.  The effective halo mass of
  MegaZ-LRG galaxies lies in the range ${\rm log}_{10} (M_{\rm eff} /
  h^{-1} M_\odot) = 13.61 \rightarrow 13.80$ (increasing with
  redshift, assuming large-scale normalization $\sigma_8 = 0.8$) for
  corresponding number densities in the range $n_g = 5.03 \rightarrow
  0.56 \times 10^{-4} \, h^3$ Mpc$^{-3}$.  Our results confirm the
  usefulness of the halo model for gaining physical insight into the
  patterns of galaxy clustering.
\end{abstract}

\section{Introduction}
\renewcommand{\thefootnote}{\fnsymbol{footnote}}
\setcounter{footnote}{1}
\footnotetext{E-mail: cblake@astro.swin.edu.au}

Photometric redshift surveys offer a route to delineating the
large-scale structure of the Universe that is increasingly competitive
with spectroscopic redshift surveys (Budavari et al.\ 2003; Seo \&
Eisenstein 2003; Amendola, Quercellini \& Giallongo 2004; Dolney, Jain
\& Takada 2004; Blake \& Bridle 2005; Phleps et al.\ 2006; Zhan et
al.\ 2006, Blake et al.\ 2007; Padmanabhan et al.\ 2007).  The ease
with which modern imaging surveys can map large areas of sky to faint
magnitude limits compensates for the absence of precise (but
time-consuming) spectroscopic redshift measurements for individual
galaxies.  An absolute pre-requisite, however, is the availability of
high-quality photometric galaxy redshifts with known error
distributions (established for example via spectroscopy of
sub-samples), together with accurate survey photometric calibration
over large angles of sky.  Recent observational efforts have enabled
both of these criteria to be satisfied.

The Sloan Digital Sky Survey (SDSS; York et al.\ 2000) has now
provided an accurately-calibrated imaging dataset over roughly a fifth
of the sky, which can be used to extract samples of galaxies in a
uniform manner.  In particular, a photometric catalogue of Luminous
Red Galaxies (LRGs) can be readily extracted using a series of
well-understood colour and magnitude cuts (Eisenstein et al.\ 2001).
Owing to their high luminosity and typical residence in the most
massive dark matter haloes, LRGs are efficient tracers of cosmic
structure across large volumes (Brown et al.\ 2003; Zehavi et
al.\ 2005a; Eisenstein et al.\ 2005b).  Moreover, these objects
provide particularly reliable photometric redshifts owing to the
strong spectral break at $\approx 4000$\AA\, in the galaxy rest frame
and the consequent rapid variation with redshift of their observed
colours in the SDSS filter system (Padmanabhan et al.\ 2005; Collister
et al.\ 2007).  Furthermore, the photometric redshift error
distribution of the LRGs can be accurately calibrated owing to the
existence of spectroscopic observations of a sub-sample as part of the
2dF-SDSS LRG and Quasar (2SLAQ) survey at the Anglo-Australian
Telescope (Cannon et al.\ 2006).

The combination of these datasets has allowed us to construct a large
catalogue of more than $10^6$ LRGs spanning the redshift range $0.4 <
z < 0.7$ with an r.m.s. photometric redshift error $\sigma_z \equiv
\sqrt{<(\delta z)^2> - <\delta z>^2} \approx 0.03 (1+z)$ where $\delta
z \equiv z_{\rm phot} - z_{\rm spec}$.  We have dubbed this catalogue
``MegaZ-LRG'' (Collister et al.\ 2007).  The database covers almost
$6{,}000$ deg$^2$, or an effective volume $\approx 2.5$ $h^{-3}$
Gpc$^3$.  We have already used this catalogue to measure the
large-scale clustering of the galaxies on linear and quasi-linear
scales via a power spectrum analysis and thereby extract measurements
of cosmological parameters (Blake et al.\ 2007; see also Padmanabhan
et al.\ 2007).  In this study we turn to the small-scale clustering
properties of the LRGs.  Our goal is to present new measurements of
the small-scale correlation function of LRGs at $z \sim 0.5$, and to
connect these measurements to the physical manner in which these LRGs
populate dark matter haloes.

It has long been known that different classes of galaxy possess
different clustering properties, in a manner connected to their
small-scale environments (for recent observational studies we refer
the reader to Norberg et al.\ 2002; Budavari et al.\ 2003; Hogg et
al.\ 2003; Zehavi et al.\ 2005b; and references therein).  For many
years these differing clustering properties could be adequately
described by fitting a simple power-law function to the two-point
galaxy correlation function on small scales (Peebles 1980).  However,
recent surveys have measured the clustering pattern accurately enough
to detect deviations from the canonical clustering power-law
(e.g.\ Hawkins et al.\ 2003; Zehavi et al.\ 2004; Zheng 2004;
Eisenstein et al.\ 2005a; Zehavi et al.\ 2005a,b; Phleps et
al.\ 2006).  These deviations provide an important insight into the
processes of galaxy formation.

The richer structure in the galaxy clustering pattern revealed by
recent surveys has been successfully interpreted in terms of the
``halo model'' (e.g.\ Seljak 2000; Peacock \& Smith 2000; Scoccimarro
et al.\ 2001; Cooray \& Sheth 2002; Berlind \& Weinberg 2002; Kravtsov
et al.\ 2004; Zehavi et al.\ 2004; Zheng 2004; Zheng et al.\ 2005;
Zehavi et al.\ 2005b; Collister \& Lahav 2005; Tinker et al.\ 2005).
In this model, the small-scale clustering of a distribution of
galaxies is linked to the underlying network of dark matter haloes,
whose properties can be measured using cosmological simulations.  A
class of galaxies is assumed to populate haloes in accordance with a
statistical ``halo occupation distribution'' as a function of the halo
mass.  The clustering then naturally separates into two components:
the distribution of galaxies within individual haloes, which dominates
on small scales ($\la 1$ Mpc), and the mutual clustering of galaxies
inhabiting separate haloes, which dominates on larger scales ($\ga 1$
Mpc).  The combination of these two terms can accurately model the
observed scale-dependent features in the small-scale clustering
pattern.

The outline of this paper is as follows: in Section \ref{secdata} we
briefly describe the ``MegaZ-LRG'' dataset.  We then present
measurements of the angular correlation function in narrow redshift
slices together with simple power-law fits in Section \ref{secmeas}.
In Section \ref{sechalo} we introduce the framework of the halo model
and in Section \ref{secparfit} we re-fit the clustering measurements
by parameterizing the halo occupation distribution of the LRGs,
comparing our results to previous work in Section \ref{secprev}.
Section \ref{secsys} investigates a range of potential systematic
photometric errors in the catalogue that may bias our results.  We
conclude in Section \ref{secconc}.

Throughout our study we assume a fixed set of large-scale cosmological
parameters: fractional matter density $\Omega_{\rm m} = 0.3$,
cosmological constant $\Omega_\Lambda = 0.7$, curvature $\Omega_{\rm
  k} = 0$, Hubble parameter $h = 0.7$, fractional baryon density
$f_{\rm b} = \Omega_{\rm b}/\Omega_{\rm m} = 0.15$, slope of the
primordial power spectrum $n_{\rm s} = 1$, and overall normalization
of the power spectrum $\sigma_8 = 0.8$.  These values are consistent
with fits to the large-scale power spectrum of the LRGs (Blake et
al.\ 2007) and to the latest measurements of the anisotropy spectrum
of the Cosmic Microwave Background (Spergel et al.\ 2007).  We note
that there is some degeneracy between the assumed cosmological
parameters and the fitted halo model parameters, and we investigate
how the best-fitting halo model parameters depend on the values of
$\sigma_8$ and $n_{\rm s}$.

\section{The data set}
\label{secdata}

We analyze angular clustering in the ``MegaZ-LRG'' galaxy database, a
photometric-redshift catalogue of Luminous Red Galaxies (LRGs) based
on the imaging dataset of the SDSS 4th Data Release.  The construction
of this catalogue is described in detail by Collister et al.\ (2007),
and cosmological-parameter fitting to the angular power spectrum is
presented by Blake et al.\ (2007).  We only provide a brief
description of the catalogue here, refering the reader to these two
papers for more information.

MegaZ-LRG contains over $10^6$ LRGs spanning the redshift range $0.4 <
z < 0.7$ with an r.m.s. redshift error $\delta z \approx 0.03 (1+z)$.
The angular selection function is described by Blake et al.\ (2007)
and encompasses 5914 deg$^2$ (the three southern SDSS stripes are
excluded).  The sample was selected from the SDSS imaging database
using a series of colour and magnitude cuts (Eisenstein et al.\ 2001;
Collister et al.\ 2007), which amount to a magnitude-limited sample of
Luminous Red Galaxies with $i$-band magnitudes $17.5 < i < 20$.
Photometric redshifts were derived for these galaxies using an
Artificial Neural Network method, ANNz (Firth, Lahav \& Somerville
2003; Collister \& Lahav 2004), constrained by a spectroscopic
sub-sample of $\approx 13{,}000$ galaxies obtained by the 2SLAQ survey
(Cannon et al.\ 2006).  In this paper we analyze a ``conservative''
version of the database in which the selection cuts applied to the
imaging database are identical to those used to produce the great
majority of the spectroscopic sub-sample (e.g.\ the faint magnitude
limit is brightened to $i = 19.8$).  Star-galaxy separation cuts were
applied both in the initial selection from the SDSS database and via
the neural network analysis.  These cuts produced a catalogue with
$644{,}903$ entries, amongst which there is a $1.5\%$ M-star
contamination (see Blake et al.\ 2007).

We used the photometric redshifts to divide the sample into four
narrow redshift slices of width $\Delta z_{\rm phot} = 0.05$ between
$z = 0.45$ and $z = 0.65$ (there are very few galaxies in the
catalogue outside this redshift range).  We then applied a luminosity
threshold in each redshift range to create a ``volume-limited'' sample
of galaxies, as assumed by the halo model.  The luminosity threshold
is given by the faint apparent magnitude limit $i=19.8$ applied at the
most distant redshift of each slice.  We calculated luminosities for
each galaxy using Luminous Red Galaxy K-corrections from the 2SLAQ
survey (Wake et al.\ 2006), not including an evolution correction.
The faint absolute $i$-band magnitude limits for each redshift slice
are then $M_i - 5 {\rm log}_{10} h = (-22.23, -22.56, -22.87,
-23.20)$.  The number of galaxies remaining in each redshift slice
after the luminosity cut was $N = (168287,118863,70229,27203)$ with
corresponding surface densities $(28.5, 20.1, 11.9, 4.6)$ deg$^{-2}$.
The spectroscopic redshift distribution of galaxies in each photo-$z$
slice (including the luminosity cut) can be deduced using the 2SLAQ
spectroscopic sub-sample.  As shown by Blake et al.\ (2007), the
spec-$z$ probability distribution for each slice is well-described by
a Gaussian function; the mean $\mu$ and standard deviation $\sigma$
for each photo-$z$ slice are listed in Table \ref{tabpowfit} (note
that these values are slightly different from those listed in Blake et
al.\ 2007, due to the additional luminosity threshold applied in the
current study).  These redshift distributions are used to project the
model correlation function to fit the observed angular clustering in
each slice.

\section{Angular correlation function measurements}
\label{secmeas}

\subsection{Method}

We used the Landy \& Szalay (1993) estimator to measure $w(\theta)$
for each photometric redshift slice in 30 logarithmically-spaced
angular separation bins between $\theta = 0.001^\circ$ and $\theta =
1^\circ$.  For each redshift slice we generated 10 random datasets
across the survey geometry, each containing the same number of
galaxies as the survey datasets.  We assume that the $1.5\%$ stellar
contamination is distributed evenly across the redshift slices in an
unclustered fashion such that the amplitude of the measured angular
correlation function is simply reduced by a constant factor $(1-f)^2$
where $f = 0.015$.  We corrected our estimates of $w(\theta)$ and
corresponding errors upwards by this factor (of $3.0\%$).

\subsection{Error determination}

\begin{figure}
\center
\epsfig{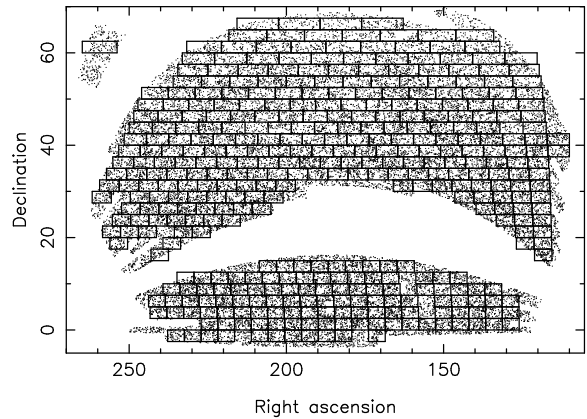}
\caption{The grid of $N=393$ sub-fields of equal area, defined by
  constant right ascension and declination boundaries, which was used
  to estimate the error in the angular correlation function via
  jack-knife re-sampling of the measurement.}
\label{figjack}
\end{figure}

We estimated the covariance matrix of the errors in the separation
bins using the technique of jack-knife resampling.  For each
measurement of $w(\theta)$ we divided the survey area into $N = 393$
sub-fields of constant area (of $13.3$ deg$^2$) using a grid of right
ascension and declination divisions (see Figure \ref{figjack}).  When
creating the grid, we allowed a sub-field to contain up to $20\%$
fractional area beyond survey edges or of survey holes.  The number of
sub-fields was chosen to be sufficient for estimating each unique
element of the covariance matrix used in the model fitting with
statistical independence.  However, we checked that our best-fitting
parameters did not depend on the number of sub-fields, or on the
restriction of our fits to the most significant principal components
of the covariance matrix using singular value decomposition
techniques.

Having defined the grid of $N$ sub-fields, we then measured
$w(\theta)$ $N$ times, which we label as $w_i(\theta)$ from $i=1$ to
$i=N$, in each case omitting just one of the sub-fields (and using the
remaining $N-1$ fields).  The covariance of the measurements between
separation bins $j$ and $k$ was deduced as:

\begin{eqnarray}
C_{jk} &\equiv& \left< w(\theta_j) \, w(\theta_k) \right> - \left<
w(\theta_j) \right> \left< w(\theta_k) \right> \\ &\approx& (N-1)
\left( \frac{\sum_i^N w_i(\theta_j) w_i(\theta_k)}{N} -
\overline{w(\theta_j)} \times \overline{w(\theta_k)} \right)
\end{eqnarray}
where $\overline{w(\theta_j)} = \sum_i^N w_i(\theta_j) / N$.

\begin{figure}
\center
\epsfig{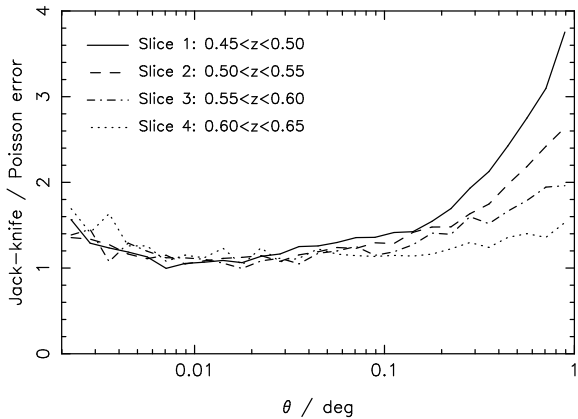}
\caption{The error in the angular correlation function determined by
  jack-knife re-sampling for the four redshift slices, normalized by
  the result assuming simple Poisson statistics.}
\label{figerr}
\end{figure}

\begin{figure}
\center
\epsfig{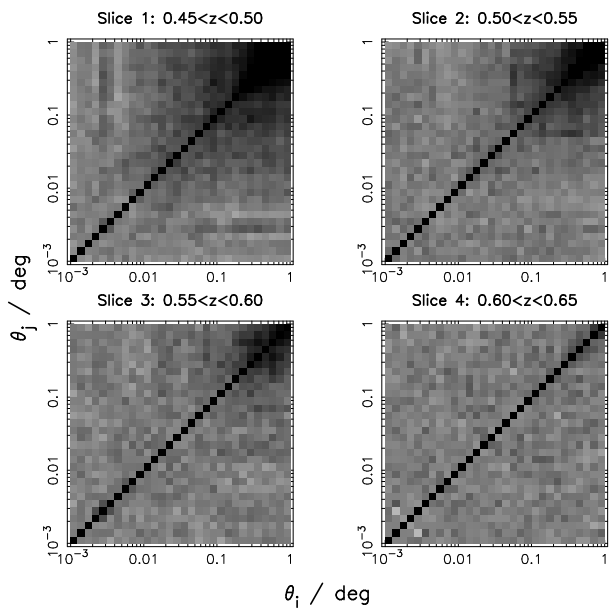}
\caption{Grey-scale plot of the correlation coefficient $r$ of
  equation \ref{eqcov}, indicating the degree of covariance between
  different separation bins for each redshift slice.  The
  correspondence of $r$ to the grey-scale is chosen as $r=-0.5$
  (white) to $r=0.5$ (black).}
\label{figcov}
\end{figure}

Figure \ref{figerr} plots the jack-knife errors in the separation bins
for each redshift slice, normalized by the error determined assuming
simple Poisson statistics (for which the error in the data pair count
of $DD$ objects in a separation bin is $\sqrt{DD}$).  For most of the
angular range under investigation, the jack-knife errors are less than
$50\%$ higher than those predicted by Poisson statistics.  For the
largest angular scales $\theta$ considered, the jack-knife error
increases relative to the Poisson error, owing to the increasing
importance of edge effects and the heightened ``cosmic variance'' owing
to the reduced number of independent cells of size $\theta$ that can
be accommodated by the dataset.  This is the familiar result from
clustering measurements that Poisson noise dominates on small scales,
and cosmic variance dominates on large scales.  The full covariance
matrices are displayed in Figure \ref{figcov} by plotting in
grey-scale the ``correlation coefficient'' between two separation bins
$i$ and $j$:
\begin{equation}
r(i,j) = \frac{{\rm Cov}(i,j)}{\sqrt{{\rm Cov}(i,i) \, {\rm Cov}(j,j)}}
\label{eqcov}
\end{equation}
Angular correlation function measurements in large separation bins are
positively correlated.  The amplitude of these correlations decreases
with redshift because, as the number density of the sample reduces,
Poisson noise becomes more important compared to cosmic variance.
These covariance matrices are always used in our model fitting.

\subsection{Power-law fits}
\label{secpow}

\begin{figure*}
\center
\epsfig{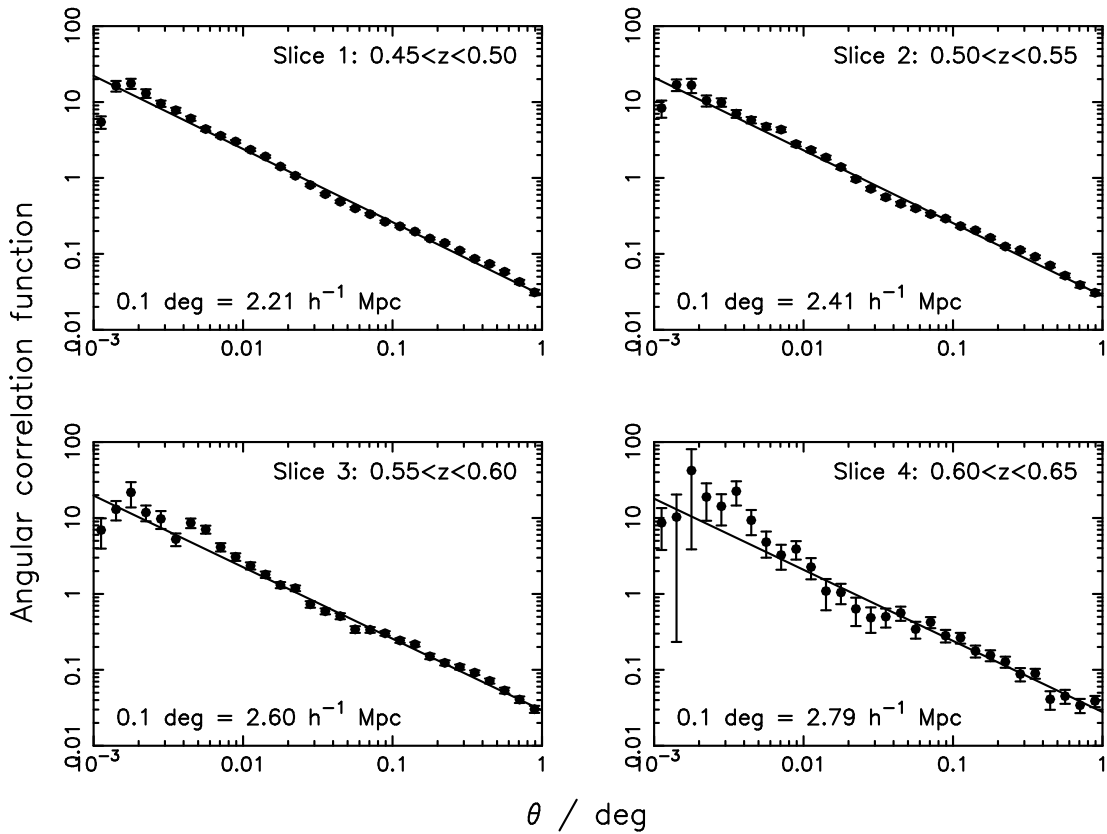}
\caption{The measured angular correlation function in the four
  redshift slices together with the best-fitting power laws.  The
  turnover of the correlation function towards very small angular
  separations is not a property of galaxy clustering; we restrict our
  fit to separations $\theta > 0.002^\circ$.  The co-moving spatial
  scale corresponding to angular separation $0.1^\circ$ at the median
  redshift of each slice is displayed in the bottom left-hand corner
  of each plot.}
\label{figcorrpowfit}
\end{figure*}

As an initial step we fitted power-laws $w(\theta) = a \,
\theta^{1-\gamma}$ to the measured angular correlation functions in
the four redshift slices (using chi-squared minimization with the full
covariance matrix).  We excluded the first 3 data points with
separations $\theta < 0.002^\circ$ ($= 7$ arcsec), as these scales are
potentially affected by astronomical seeing and issues of galaxy
merging and blending (the decrease in the value of the correlation
functions at the smallest scales in Figure \ref{figcorrpowfit} is not
a property of galaxy clustering).  The best-fitting power-law
parameters $(a,\gamma)$ are listed in Table \ref{tabpowfit} and the
models and data points are displayed in Figure \ref{figcorrpowfit}.
The best-fitting slopes $\gamma$ for the redshift slices take values
in the range $1.94 \rightarrow 1.96$, consistent with previous studies
of Luminous Red Galaxies (e.g.\ Eisenstein et al.\ 2005a; Zehavi et
al.\ 2005a).  However, as indicated by the high values of the minimum
chi-squared statistic $\chi^2_{\rm pl}$ in Table \ref{tabpowfit}
compared to the 25 degrees of freedom, a power-law is not a good fit
to the data.  The fact that we detect deviations from simple power-law
clustering with high significance indicates that our photo-$z$ survey
can be used to fit more complex and physically-insightful models to
the small-scale clustering pattern.

\begin{table*}
\center
\caption{Power-law fits to the angular correlation function in the
  four redshift slices.  Column 2 records the surface density of the
  galaxy sample in each redshift slice.  Column 3 is the threshold
  galaxy absolute $i$-band magnitude in each slice for the
  volume-limited sample.  Columns 4 and 5 list the parameters of the
  Gaussian redshift distribution for each slice, $p(z) \propto
  \exp{[-(z-\mu)^2/2\sigma^2]}$.  Columns 6 and 7 show the
  best-fitting power-law parameters $w(\theta) = a \,
  \theta^{1-\gamma}$, where $\theta$ is in degrees.  The associated
  errors for $a$ and $\gamma$ were obtained by marginalizing over
  values of the other parameter.  Note that the amplitudes have been
  corrected upwards by a stellar contamination factor $(1-f)^{-2} =
  1.03$.  Column 8 is the corresponding chi-squared statistic
  $\chi^2_{\rm pl}$ evaluated using the full covariance matrix (which
  we compare with 25 degrees of freedom).  Columns 9 lists the
  inferred galaxy clustering length $r_0$ of the spatial correlation
  function $\xi(r) = (r/r_0)^{-\gamma}$.}
\begin{tabular}{ccccccccc}
\hline

Redshift slice & Density & $M_i - 5$ log$_{10}h$ & $\mu$ & $\sigma$ &
$10^2 \times a$ & $\gamma$ & $\chi^2_{\rm pl}$ & $r_0$ \\

& (deg$^{-2}$) & & & & (deg$^{\gamma-1}$) & & & ($h^{-1}$ Mpc) \\
\hline
$0.45 < z < 0.5$ & 28.5 & $-22.23$ & 0.476 & 0.034 & $2.84 \pm 0.12$ & $1.96 \pm 0.01$ & 177.1 & $7.2 \pm 0.2$ \\
$0.5 < z < 0.55$ & 20.1 & $-22.56$ & 0.528 & 0.040 & $2.79 \pm 0.14$ & $1.96 \pm 0.01$ & 132.8 & $8.0 \pm 0.2$ \\
$0.55 < z < 0.6$ & 11.9 & $-22.87$ & 0.574 & 0.043 & $2.93 \pm 0.18$ & $1.94 \pm 0.02$ & 71.8 & $8.5 \pm 0.3$ \\
$0.6 < z < 0.65$ & 4.6 & $-23.20$ & 0.629 & 0.052 & $2.78 \pm 0.19$ & $1.94 \pm 0.02$ & 34.9 & $9.3 \pm 0.3$ \\
\hline
\end{tabular}
\label{tabpowfit}
\end{table*}

Nevertheless, we can use the power-law amplitude of the angular
clustering to estimate the three-dimensional clustering length $r_0$
of the population of galaxies, using the known redshift distribution
$p(z)$ in each slice.  If we define the spatial two-point correlation
function $\xi(r) = (r/r_0)^{-\gamma}$ then the amplitude of the
power-law angular correlation function $w(\theta) = a \,
\theta^{1-\gamma}$ follows from Limber's equation:
\begin{equation}
a = C_\gamma \, r_0^\gamma \int dz \, p(z)^2 \left( \frac{dx}{dz}
\right)^{-1} x(z)^{1-\gamma}
\end{equation}
where $x(z)$ is the co-moving radial co-ordinate at redshift $z$ and
$C_\gamma = \Gamma(\frac{1}{2}) \Gamma(\frac{\gamma}{2} - \frac{1}{2})
/ \Gamma(\frac{\gamma}{2})$.  In Table \ref{tabpowfit} we determine
the corresponding values of $r_0$ and $\gamma$ for each redshift
slice.  The amplitude of the clustering length, $r_0 = 7.2 \rightarrow
9.3$ $h^{-1}$ Mpc, is consistent with highly-biased massive galaxies.
Our small-scale angular clustering measurements are not affected by
redshift-space distortions, since the photo-$z$ errors are much bigger
than the peculiar velocities of the galaxies.  Correlated
redshift-space distortions are important on larger scales however, as
discussed in Blake et al.\ (2007) and Padmanabhan et al.\ (2007).

The clustering amplitude systematically increases with redshift for
two reasons:
\begin{enumerate}
\item Each volume-limited sample in successive redshift slices has the
  same limiting apparent magnitude hence higher luminosity threshold.
  The higher-redshift galaxies are hence preferentially more luminous
  and more strongly clustered (e.g.\ Norberg et al.\ 2002; Zehavi et
  al.\ 2005b).  In Table \ref{tabpowfit} we list the threshold
  absolute $i$-band magnitude $M_i$ of galaxies in each redshift
  slice, calculated using a Luminous Red Galaxy K-correction (Wake et
  al.\ 2006).
\item In standard models of the evolution of galaxy clustering, the
  bias factor of a class of galaxies increases with redshift in
  opposition to the decreasing linear growth factor, in order to
  reproduce the observed approximate constancy of the small-scale
  co-moving clustering length (e.g.\ Lahav et al.\ 2002).  Simple
  models for this effect such as $b(z) = 1 + (b_0 - 1)/D(z)$ (Fry
  1996) or $b(z) = b_0/D(z)$, where $D(z)$ is the linear growth
  factor, predict an evolution in bias across our analyzed redshift
  range of $\Delta b \approx 0.2$.
\end{enumerate}
These trends are in good agreement with our measurements of the
amplitude of the large-scale clustering pattern (Blake et al.\ 2007).

\section{Halo model framework}
\label{sechalo}

We used the halo model of galaxy clustering to produce model spatial
correlation functions $\xi(r)$ to fit to our measurements.  We
summarize the ingredients of our model here.  Further details can be
found in e.g.\ Seljak 2000; Cooray \& Sheth 2002; Berlind \& Weinberg
2002; Kravtsov et al.\ 2004; Zehavi et al.\ 2004; Zheng 2004; Zehavi
et al.\ 2005b; Tinker et al.\ 2005.

In the halo model framework, the clustering functions are expressed as
a sum of components due to pairs of galaxies within a single dark
matter halo (the ``1-halo term'' $\xi_1$) and to pairs of galaxies
inhabiting separate haloes (the ``2-halo term'' $\xi_2$):
\begin{equation}
\xi(r) = [1 + \xi_1(r)] + \xi_2(r)
\end{equation}
where the ``1+'' at the start of the expression arises because the
total number of galaxy pairs ($\propto 1 + \xi$) is the sum of the
number of pairs from single haloes ($\propto 1 + \xi_1$) and from
different haloes ($\propto 1 + \xi_2$).  The two terms dominate on
different scales, with the 1-halo term only important on small scales
$\la 1$ Mpc.

The fundamental ingredient of the galaxy halo model is the {\it halo
  occupation distribution} (HOD), which describes the probability
distribution for the number of galaxies $N$ hosted by a dark matter
halo as a function of its mass $M$.  In order to construct the 1-halo
and 2-halo two-point clustering terms, we require the first and second
factorial moments of the HOD, $<N|M>$ and $<N(N-1)|M>$.  We make the
assumption that the first galaxy to be hosted by a halo lies at the
centre of the halo, and any remaining galaxies are classified as
``satellites'' and distributed in proportion to the halo mass profile.
We apply different HODs for the central and satellite galaxies,
$<N_c|M>$ and $<N_s|M>$ respectively, where
\begin{equation}
<N|M> = <N_c|M> (1 + <N_s|M>)
\label{eqnm}
\end{equation}
Equation \ref{eqnm} takes its form because a halo can only host a
satellite galaxy if it already contains a central galaxy.  We will use
the notation $N_c(M) \equiv <N_c|M>$, $N_s(M) \equiv <N_s|M>$ and
$N(M) \equiv <N|M>$ in the equations that follow.

\subsection{The 1-halo term $\xi_1(r)$}

The 1-halo galaxy correlation function is composed of contributions
from central-satellite pairs and satellite-satellite pairs.  It is
convenient to evaluate these two contributions separately.  The 1-halo
correlation function for central-satellite pairs is given by:
\begin{equation}
  1 + \xi_{1,c-s}(r) = \int_{M_{\rm vir}(r)}^\infty dM \, n(M)
  \frac{N_c(M) N_s(M)}{n_g^2/2} \frac{\rho(r|M)}{M}
\end{equation}
where $n_g$ is the galaxy number density, $n(M)$ is the halo mass
function, and $\rho(r|M)$ is the halo density profile.  The lower
limit for the integral is the halo mass $M$ corresponding to a virial
radius $r$, given that less massive haloes have smaller radii and
cannot contribute any central-satellite galaxy pairs with co-moving
separation $r$:
\begin{equation}
M_{\rm vir}(r) = \frac{4}{3} \pi r^3 \overline{\rho} \Delta
\end{equation}
where $\overline{\rho} = 2.78 \times 10^{11} \Omega_{\rm m}$ $h^2
M_\odot$ Mpc$^{-3}$ is the co-moving background density of the
Universe, and $\Delta = 200$ is the critical overdensity for
virialization.

It is simplest to evaluate the 1-halo correlation function for
satellite-satellite pairs in Fourier space (where convolutions become
multiplications).  The power spectrum is:
\begin{equation}
  P_{1,s-s}(k) = \int_0^\infty dM \, n(M) \frac{N_c(M) N_s^2(M)}{n_g^2} |u(k|M)|^2
\label{eq1haloss}
\end{equation}
where $u(k|M)$ is the Fourier transform of the halo density profile
$\rho(r|M)$.  Because satellite galaxies are Poisson-distributed, we
can write $<N_s (N_s - 1)> = <N_s>^2$ to obtain the above equation.
The correlation function corresponding to equation \ref{eq1haloss} is
then
\begin{equation}
\xi_{1,s-s}(r) = \frac{1}{2\pi^2} \int_0^\infty dk \, P_{1,s-s}(k) \,
k^2 \, \frac{\sin{kr}}{kr}
\end{equation}
The total 1-halo correlation function is then derived as
\begin{equation}
\xi_1 = \xi_{1,c-s} + \xi_{1,s-s}
\end{equation}

\subsection{The 2-halo term $\xi_2(r)$}

The 2-halo galaxy correlation function at separation $r$ is evaluated
from the scale-dependent 2-halo power spectrum $P_2(k,r)$:
\begin{eqnarray}
  \lefteqn{P_2(k,r) = P_m(k) \times} \nonumber \\ & & \left[
    \int_0^{M_{\rm lim}(r)} dM \, n(M) \, b(M,r) \frac{N(M)}{n_g'(r)}
    u(k|M) \right]^2
\label{eq2halo}
\end{eqnarray}
where $P_m(k)$ is the non-linear matter power spectrum at the survey
redshift, $b(M,r)$ is the scale-dependent halo bias at separation $r$,
and $n_g'(r)$ is the restricted galaxy number density at separation
$r$, where
\begin{equation}
n_g'(r) = \int_0^{M_{\rm lim}(r)} dM \, n(M) \, N(M)
\label{eqdens2}
\end{equation}
The mass truncation $M_{\rm lim}(r)$ must be included to incorporate
the effects of {\it halo exclusion}: more massive haloes would overlap
at separation $r$.  We derive the mass limit using the
``$n_g'$-matched'' approximation described in Tinker et al.\ (2005).
Firstly we calculate the restricted number density using equation B13
in Tinker et al., which includes the effects of triaxiality:
\begin{eqnarray}
  \lefteqn{n_g'(r) = \int_0^\infty dM_1 \, n(M_1) N(M_1) \times} \nonumber \\ & & \int_0^\infty dM_2 \, n(M_2) \, N(M_2) \, P(r,M_1,M_2)
\end{eqnarray}
where $P(r,M_1,M_2)$ quantifies the probability of non-overlapping
haloes of masses $M_1$ and $M_2$ with separation $r$.  Defining $x =
r/(R_1 + R_2)$, where $R_1$ and $R_2$ are the virial radii
corresponding to masses $M_1$ and $M_2$, and using $y = (x-0.8)/0.29$,
then Tinker et al.\ obtain $P(y) = 3y^2 - 2y^3$ from simulations.
Given this value of $n_g'(r)$, we increase the value of $M_{\rm
  lim}(r)$ in equation \ref{eqdens2} to produce a matching number
density.  This value is then used in equation \ref{eq2halo} to produce
the 2-halo power spectrum $P_2(k,r)$.

Following Tinker et al.\ (2005), we assumed the following model for
the scale-dependent bias:
\begin{equation}
b^2(M,r) = b^2(M) \frac{\left[ 1 + 1.17 \, \xi_m(r)
    \right]^{1.49}}{\left[ 1 + 0.69 \, \xi_m(r) \right]^{2.09}}
\end{equation}
where $\xi_m(r)$ is the non-linear matter correlation function.  We
derived matter power spectra using the ``{\tt CAMB}'' software package
(Lewis, Challinor \& Lasenby 2000), including corrections for
non-linear growth of structure using the fitting formulae of Smith et
al.\ (2003) (``{\tt halofit=1}'' in {\tt CAMB}).  We outputted power
spectra at the mean redshift of each slice, and obtained the
correlation function using a Fourier transform.  The 2-halo galaxy
correlation function is obtained via the Fourier transform of equation
\ref{eq2halo}:
\begin{equation}
\xi_2'(r) = \frac{1}{2\pi^2} \int_0^\infty dk \, P_2(k,r) \, k^2 \,
\frac{\sin{kr}}{kr}
\end{equation}
with the number of galaxy pairs corrected from the restricted galaxy
density to the entire galaxy population:
\begin{equation}
1 + \xi_2(r) = \left[ \frac{n'_g(r)}{n_g} \right]^2 [1 + \xi_2'(r)]
\end{equation}

\subsection{Halo mass and bias functions}

The {\it halo mass function} $n(M)$ describes the number density of
haloes as a function of mass $M$.  We introduce the new mass variable
$\nu$ following Press \& Schechter (1974):
\begin{equation}
n(M) \, dM = \frac{\overline{\rho}}{M} \, f(\nu) \, d\nu
\end{equation}
The new mass variable $\nu$ is defined by $\nu \equiv \left[
  \delta_{\rm sc} / \sigma(M,z) \right]^2$, where $\delta_{\rm sc}$ is
the linear-theory prediction for the present-day overdensity of a
region which would undergo spherical collapse at redshift $z$, and
$\sigma^2(M,z)$ is the variance of the linear power spectrum in a
spherical top hat which contains average mass $M$:
\begin{equation}
  \sigma^2(M,z) = \frac{D(z)^2}{2 \pi^2} \int_0^\infty dk \, k^2 \, P_{\rm lin}(k) \, W^2(kR)
\label{eqsigsq}
\end{equation}
where $W(x) = (3/x^3) [\sin{x} - x \cos{x}]$, $M = \frac{4}{3} \pi R^3
\overline{\rho}$, $D(z)$ is the linear growth factor at redshift $z$,
and $P_{\rm lin}(k)$ is the linear power spectrum at redshift zero.
We approximate $\delta_{\rm sc} = 1.686$ independently of redshift.

We use the Jenkins et al.\ (2001) model for the mass function:
\begin{equation}
  \nu f(\nu) = \frac{1}{2} a_1 \exp{[-|\ln{(\sqrt{\nu}/\delta_{\rm sc})} + a_2|^{a_3}]}
\end{equation}
where $a_1 = 0.315$, $a_2 = 0.61$ and $a_3 = 3.8$.

The {\it halo bias function} $b(M)$ describes the biasing of a halo of
mass $M$ with respect to the overall dark matter distribution.  We use
the Sheth, Mo \& Tormen (2001) model for the bias function, with the
revised parameters stated in Tinker at al.\ (2005):
\begin{eqnarray}
\lefteqn{b(\nu) = 1 + \frac{1}{\delta_{\rm sc}} \times} \nonumber
\\ & & \left[ q\nu + s (q \nu)^{1-t} - \frac{q^{-1/2}}{1 +
    s(1-t)(1-\frac{t}{2})(q\nu)^{-t}} \right]
\end{eqnarray}
where the constants $q = 0.707$, $s = 0.35$ and $t = 0.8$.

\subsection{Halo profiles}

We use the Navarro, Frenk \& White (1997) dark matter halo density
profile:
\begin{equation}
\rho(r) = \frac{\rho_s}{(r/r_s)(1 + r/r_s)^2} \hspace{5mm} (r < r_{\rm
  vir})
\end{equation}
where $r_s$ is the characteristic scale radius and $\rho_s$ provides
the normalization.  The profile is truncated at the virial radius
$r_{\rm vir}$, which is obtained from the halo mass via
\begin{equation}
r_{\rm vir} = \left( \frac{3M}{4\pi\Delta\overline{\rho}}
  \right)^{1/3}
\end{equation}
We parameterize the profile in terms of the {\it concentration
  parameter} $c = r_{\rm vir}/r_s$.  The normalization for the mass
$M$ is
\begin{equation}
M = \int_0^{r_{\rm vir}} \rho(r) 4\pi r^2 dr = 4 \pi \rho_s r_s^3
\left[ \ln{(1+c)} - \frac{c}{1+c} \right]
\end{equation}
We assume that the concentration parameter $c$ depends on halo mass
$M$ and redshift $z$ in a manner calibrated by numerical simulations
(Bullock et al.\ 2001; Zehavi et al.\ 2004):
\begin{equation}
c(M,z) = \frac{11}{1+z} \left( \frac{M}{M_0} \right)^{-0.13}
\end{equation}
where $M_0$ is obtained from equation \ref{eqsigsq} by setting
$\sigma(M_0,0) = \delta_{\rm sc}$.  For our adopted cosmological model
we obtain $M_0 = 12.64 \, h^{-1} M_\odot$, resulting in a
concentration $c = 5$ for a halo of mass $M = 10^{14} \, h^{-1}
M_\odot$ at $z = 0.5$.  This assumption is consistent with the
measurement by Mandelbaum et al.\ (2006b) of the concentration
parameter for LRGs using galaxy-galaxy weak lensing.  Fitting the
concentration $c$ as an extra free parameter produced no improvement
in the minimum value of $\chi^2$.  In fact, there is a significant
degeneracy between $c$ and $\beta$.

\subsection{Halo occupation distribution}
\label{sechod}

The halo occupation distribution (HOD) is a parameterized description
of how galaxies populate dark matter haloes as a function of the halo
mass $M$.  As described above, we separate the distribution into
separate HODs for central galaxies and for satellite galaxies.  We
adopt simple models motivated by results from simulations and
semi-analytic calculations (e.g.\ Kauffmann, Nusser \& Steinmetz 1997;
Benson et al.\ 2000; Berlind et al.\ 2003; Kravtsov et al.\ 2004).

For central galaxies, our basic model is a step function such that
haloes above a minimum mass threshold $M_{\rm cut}$ contain a single
central galaxy and haloes below this threshold contain no galaxies.
However, our observational data consists of galaxy luminosities (with
a scatter resulting from the photo-$z$ errors) rather than galaxy
masses.  We therefore follow Zheng et al.\ (2005) and ``soften'' the
transition from zero to one galaxies with a further parameter
$\sigma_{\rm cut}$ such that
\begin{equation}
<N_c|M> = 0.5 \left[ 1 + {\rm erf} \left( \frac{{\rm
      log}_{10}(M/M_{\rm cut})}{\sigma_{\rm cut}} \right) \right]
\label{eqhod1}
\end{equation}
The scatter in halo mass for a fixed galaxy luminosity results from
galaxy formation physics and (in the case of this study) from
photo-$z$ errors.

For satellite galaxies, a power-law in mass provides a good
description for the mean occupation number in simulations:
\begin{equation}
<N_s|M> = \left( \frac{M}{M_0} \right)^\beta
\label{eqhod2}
\end{equation}
Introducing a ``cut-off'' to the satellite HOD of the form
\begin{equation}
<N_s|M> = \left( \frac{M-M_1}{M_0} \right)^\beta
\end{equation}
(Zheng et al.\ 2005) did not significantly improve the fit of the
model to the data.  Our halo model power spectrum is hence specified
by four variables: $M_{\rm cut}$, $\sigma_{\rm cut}$, $M_0$ and
$\beta$.

\begin{figure}
\center
\epsfig{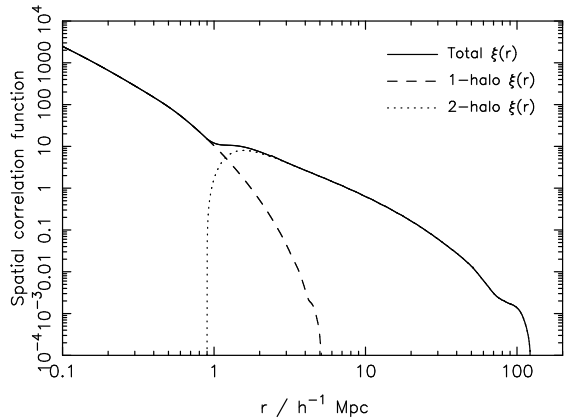}
\caption{An example halo model galaxy correlation function at $z =
  0.5$ for parameters $\sigma_{\rm cut} = 0.3$, ${\rm log}_{10} ( M_0
  / h^{-1} M_\odot) = 14.3$ and $\beta = 1.5$.  Matching galaxy
  density $n_g = 5 \times 10^{-4} \, h^3$ Mpc$^{-3}$ requires ${\rm
    log}_{10} ( M_{\rm cut} / h^{-1} M_\odot) = 12.98$.  The separate
  1-halo and 2-halo components are shown, which cross-over in
  dominance at scale $\sim 1 h^{-1}$ Mpc.  This model has a linear
  bias $b_g = 1.82$ and an effective halo mass ${\rm log}_{10} (
  M_{\rm eff} / h^{-1} M_\odot) = 13.48$.}
\label{figxihalo}
\end{figure}

In addition we must satisfy one extra constraint: the galaxy number
density
\begin{equation}
n_g = \int_0^\infty dM \, n(M) \, N(M)
\end{equation}
must match the observed number density in each redshift slice.  We
match this constraint by fixing the variable $M_{\rm cut}$ for each
choice of $\sigma_{\rm cut}$, $M_0$ and $\beta$.  Our model hence
contains three independent parameters.  Figure \ref{figxihalo}
displays an example halo model correlation function and the component
1-halo and 2-halo terms.

Useful quantities which can be derived from the HOD are the effective
large-scale bias
\begin{equation}
b_g = \int dM \, n(M) \, b(M) \frac{N(M)}{n_g}
\label{eqbg}
\end{equation}
and the effective mass $M_{\rm eff}$ of the halo occupation
distribution:
\begin{equation}
M_{\rm eff} = \int dM \, M \, n(M) \frac{N(M)}{n_g}
\label{eqmeff}
\end{equation}
We can also determine the average fraction of central or satellite
galaxies in the sample, e.g.\ for central galaxies:
\begin{equation}
  f_c = \frac{\int dM \, n(M) \, N_c(M)}{\int dM \, n(M) \, N_c(M) \, [1 +
    N_s(M)]}
\end{equation}
and $f_s = 1 - f_c$ for satellite galaxies.

\subsection{Conversion to angular correlation function}

Knowing the redshift distribution $p(z)$ of a galaxy population, we
can project the spatial galaxy correlation function $\xi(r)$ to an
angular correlation function $w(\theta)$ using Limber's equation:
\begin{equation}
w(\theta) = 2 \int_0^\infty dx \, f(x)^2 \int_0^\infty du \; \xi(r=\sqrt{u^2 + x^2 \theta^2})
\label{eqwth1}
\end{equation}
where $f(x)$ describes the radial distribution of sources as:
\begin{equation}
f(x) = \frac{p(z)}{dx/dz(z)}
\label{eqwth2}
\end{equation}
where $z$ is the redshift corresponding to co-moving radial
co-ordinate $x(z)$.

\section{Parameter fits}
\label{secparfit}

We fitted the 3-parameter halo model $(\sigma_{\rm cut}, M_0, \beta)$
to the observed angular correlation functions in each redshift slice,
fixing the remaining parameter $M_{\rm cut}$ by matching to the
observed galaxy number density $n_g$.  We used a combination of a
coarse grid-based search and a downhill-simplex method to locate the
minimum value of the $\chi^2$ statistic, using the full covariance
matrix.  We then employed a fine grid-based method to explore the
$\chi^2$ surface around the minimum and determine the errors in the
fitted parameters (by marginalizing over the other model parameters).
The mean and standard deviation of each model parameter, marginalizing
over the other parameters, are listed in Table \ref{tabhalofit}.

\begin{figure*}
\center
\epsfig{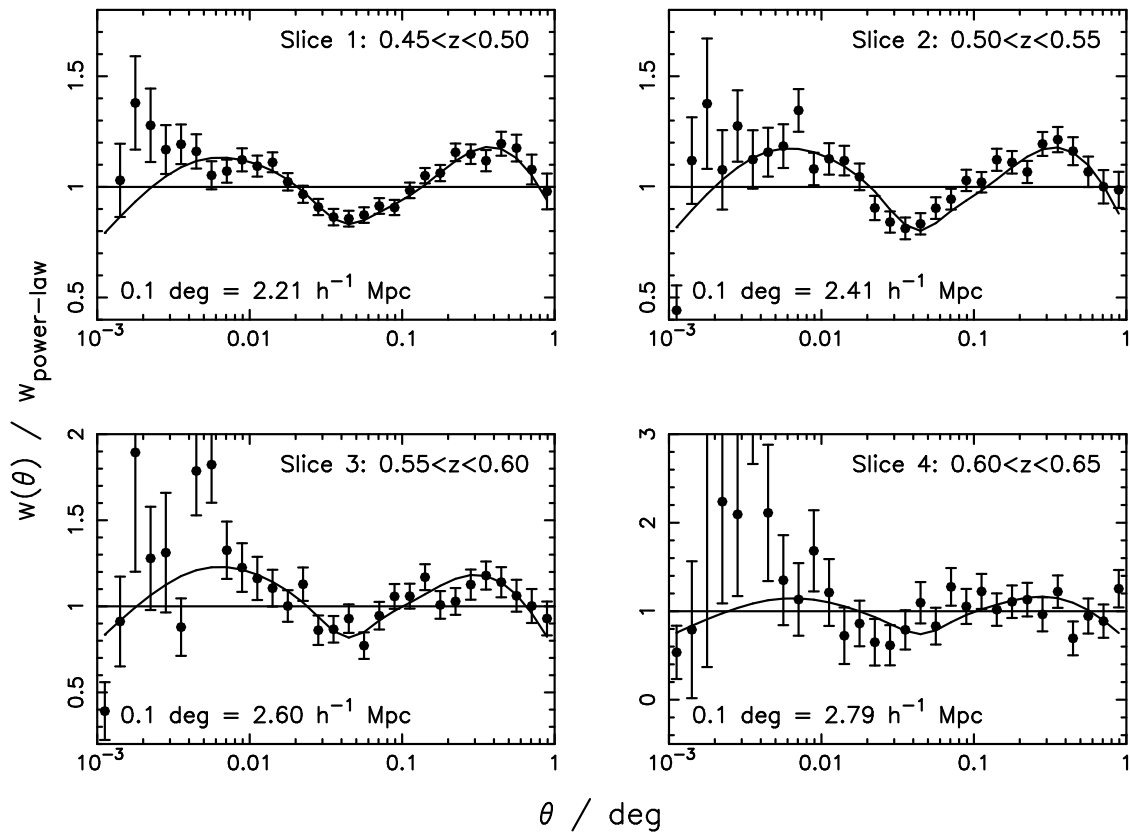}
\caption{The measured angular correlation functions in the four
  redshift slices together with the best-fitting halo models.  The
  $y$-axis is the ratio of the angular correlation function
  $w(\theta)$ and the best-fitting power-law model in each slice (from
  Section \ref{secpow}).  The co-moving spatial scale corresponding to
  angular separation $0.1^\circ$ at the median redshift of each slice
  is displayed in the bottom left-hand corner of each plot.  A flat
  $\Lambda$CDM Universe is assumed with cosmological parameters
  $\Omega_{\rm m} = 0.3$, $h = 0.7$, $\Omega_{\rm b}/\Omega_{\rm m} =
  0.15$, $\sigma_8 = 0.8$ and $n_{\rm s} = 1$.}
\label{figcorrnorm}
\end{figure*}

In Figure \ref{figcorrnorm} we plot the best-fitting halo model
correlation functions together with the data.  We divide the results
by the best-fitting power-law model from Section \ref{secpow} for
increased clarity.  We note immediately that the halo model framework
has successfully reproduced the deviations from the power-law, owing
to the separate contributions of the 1-halo and 2-halo terms.  The
result is a good fit of model to data, with the addition of only one
extra parameter compared to the original power-law fit.  The minimum
values of $\chi^2$ are around 30 (for 24 degrees of freedom).
Inspection of Figure \ref{figcorrnorm} reveals that the best-fitting
model always describes the transition region between the 1-halo and
2-halo terms very well, and that the main source of discrepancy is at
very small scales ($< 200$ kpc) where the data points lie
systematically above the model prediction.

\begin{table*}
\center
\caption{Halo model fits to the angular correlation function in the
  four redshift slices.  Column 2 records the observed galaxy number
  density which our models are constrained to match.  Columns 3, 4 and
  5 list the best-fitting values and errors of the halo model
  parameters $\sigma_{\rm cut}$, $M_0$ and $\beta$ defined by
  equations \ref{eqhod1} and \ref{eqhod2}.  Column 6 displays the
  inferred value of the final halo model parameter $M_{\rm cut}$.  In
  Columns 7 and 8 we evaluate the corresponding galaxy bias factor
  $b_g$ (equation \ref{eqbg}) and effective halo mass $M_{\rm eff}$
  (equation \ref{eqmeff}).  Column 9 records the minimum value of the
  chi-squared statistic for the halo model, $\chi^2_{\rm halo}$,
  evaluated using the full covariance matrix (which we compare with 24
  degrees of freedom).  Base-10 logarithms are used in this Table.}
\begin{tabular}{ccccccccc}
\hline

Redshift slice & $10^4 \times n_g$ & $\sigma_{\rm cut}$ & ${\rm log}
\left( \frac{M_0}{M_\odot/h} \right)$ & $\beta$ & ${\rm log} \left(
\frac{M_{\rm cut}}{M_\odot/h} \right)$ & $b_g$ & ${\rm log} \left(
\frac{M_{\rm eff}}{M_\odot/h} \right)$ & $\chi^2_{\rm halo}$ \\ &
($h^3$ Mpc$^{-3}$) & & & & & & & \\

\hline

$0.45 < z < 0.5$ & $5.03$ & $0.21 \pm 0.11$ & $14.09 \pm 0.01$ & $1.57
\pm 0.02$ & $12.98$ & $1.92 \pm 0.02$ & $13.61 \pm 0.01$ & 22.1 \\

$0.5 < z < 0.55$ & $3.07$ & $0.07 \pm 0.07$ & $14.22 \pm 0.01$ & $1.69
\pm 0.04$ & $13.12$ & $2.15 \pm 0.01$ & $13.67 \pm 0.01$ & 28.3 \\

$0.55 < z < 0.6$ & $1.60$ & $0.24 \pm 0.12$ & $14.39 \pm 0.01$ & $1.87
\pm 0.07$ & $13.35$ & $2.38 \pm 0.04$ & $13.74 \pm 0.01$ & 36.8 \\

$0.6 < z < 0.65$ & $0.56$ & $0.53 \pm 0.14$ & $14.76 \pm 0.10$ & $1.80
\pm 0.36$ & $13.79$ & $2.62 \pm 0.09$ & $13.80 \pm 0.03$ & 40.1 \\

\hline
\end{tabular}
\label{tabhalofit}
\end{table*}

In Table \ref{tabhalofit} we also list the derived values of the
galaxy bias factor $b_g$ and effective halo mass $M_{\rm eff}$ for
each redshift slice (calculated using equations \ref{eqbg} and
\ref{eqmeff}).  The errors in these quantities were obtained by
evaluating their values over the fine grid of halo model parameters
and weighting by the appropriate probability for the fit of model to
data at each grid point.  The range of linear bias, increasing with
redshift from $b_g \approx 1.92$ (at $z = 0.475$) to $b_g \approx
2.62$ (at $z = 0.625$), agrees well with fits to the large-scale power
spectrum of the LRGs (Blake et al.\ 2007), allowing for the differing
values of the normalization $\sigma_8$ (in this study we have assumed
$\sigma_8 = 0.8$, whereas the bias values in Blake et al.\ (2007) are
quoted for $\sigma_8 = 1$).  The effective mass, ranging from $M_{\rm
  eff} = 10^{13.61 \rightarrow 13.80} \, h^{-1} \, M_\odot$ over the
same redshift range, confirms that LRGs are hosted by massive dark
matter haloes and are highly biased tracers of the clustering pattern.
Our values for the effective mass are in the same range as the LRG
halo mass measured by Mandelbaum et al.\ (2006b) using galaxy-galaxy
lensing ($M = 10^{13.83} \, h^{-1} \, M_\odot$ for the bright sample
of Mandelbaum et al.).  We note that the systematic increase in galaxy
bias in each redshift slice is driven more by the increasing
luminosity threshold rather than redshift evolution.

\begin{figure}
\center
\epsfig{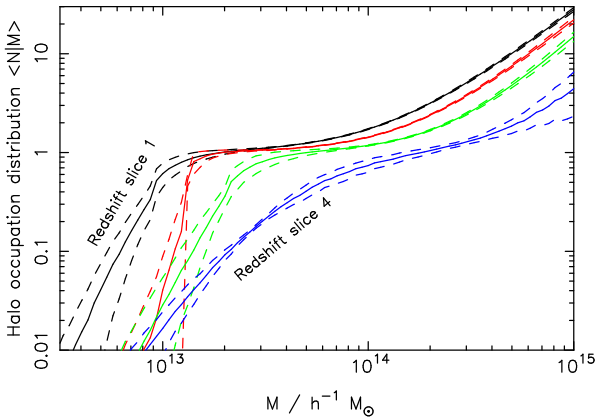}
\caption{The range of halo occupation distributions fitted to each of
  the four redshift slices (with each subsequent slice shifting from
  left to right in the Figure) assuming the parametric description of
  equations \ref{eqhod1} and \ref{eqhod2}.  This Figure was generated
  by varying the parameters $\sigma_{\rm cut}$, $M_0$ and $\beta$ over
  a grid and determining the relative probability at each grid point
  using the $\chi^2$ statistic.  At each halo mass $M$, the
  probability distribution within the model parameter space was used
  to construct the mean HOD (the solid line) and the $68\%$ confidence
  region (the dotted lines).}
\label{fighod}
\end{figure}

Figure \ref{fighod} plots the statistical range of allowed halo
occupation distributions $<N|M>$ in each of the four redshift slices
assuming the parametric description of equations \ref{eqhod1} and
\ref{eqhod2}.  The parameters $\sigma_{\rm cut}$, $M_0$ and $\beta$
were varied over a grid and the probability determined at each grid
point using the $\chi^2$ statistic.  This probability distribution was
used to construct the mean and standard deviation of the value of
$<N|M>$ as a function of halo mass $M$.  The average number of our
galaxy sample hosted by a halo of mass $M = 10^{14.5} \, h^{-1} \,
M_\odot$ is $(5.5, 4.1, 2.6, 1.4)$ in the four redshift slices,
decreasing systematically with redshift as the threshold luminosity
increases.  Broadly speaking, the effect of increasing luminosity is
to shift the HOD to higher masses without significantly changing its
shape, i.e.\ shifting to the right in Figure \ref{fighod} in a similar
fashion for central and satellite galaxies.  If we weight the HOD by
the mass function of haloes which steeply decreases with increasing
mass, we find that the fraction of galaxies that are classified as
central galaxies is very high: $f_c = (0.88, 0.90, 0.93, 0.97)$ in the
four redshift slices.  This agrees with the standard picture of
Luminous Red Galaxies forming at the heart of massive dark matter
haloes.

The best-fitting value of $\beta$, the slope of the power-law HOD for
satellite galaxies, increases systematically with redshift.  The most
important cause of this trend is evolution of $\beta$ with luminosity,
as measured at low redshift (Zehavi et al.\ 2005b).  Investigating
this further, we re-fitted the halo model parameters in each of the
redshift slices for each of the absolute $i$-band magnitude thresholds
$M_i - 5 {\rm log}_{10} h = (-22.23, -22.56, -22.87, -23.20)$ which
correspond to the luminosity thresholds of the four redshift slices.
Hence all redshift slices can provide a sample with $M_i - 5 {\rm
  log}_{10} h < -23.20$, but only the first redshift slice contributes
a sample with $M_i - 5 {\rm log}_{10} h < -22.23$.  We obtained the
probability distribution for $\beta$ for each halo model fit,
marginalizing over the other model parameters, and combined the
results for matched luminosity samples in different redshift slices.
We find that the measurements of $\beta$ at a fixed luminosity
threshold agree well between the redshift slices, and the combined
result for the four luminosity thresholds listed above is $\beta =
(1.57 \pm 0.02, 1.68 \pm 0.03, 1.91 \pm 0.05, 2.15 \pm 0.14)$,
confirming a significant evolution of $\beta$ with luminosity.
Considering just the first redshift slice, the best-fitting effective
halo mass for the four luminosity slices is ${\rm log}_{10} (M_{\rm
  eff} / h^{-1} M_\odot) = (13.61, 13.73, 13.88, 14.02)$ corresponding
to number densities $n_g = (5.03, 2.45, 1.02, 0.32) \times 10^{-4} \,
h^3$ Mpc$^{-3}$.

Our results have a significant dependence on the assumed value of
$\sigma_8 = 0.8$, which sets the overall normalization of the matter
power spectrum.  For $\sigma_8 = 0.7$, the best-fitting effective halo
mass in each of the four redshift slices is ${\rm log}_{10} (M_{\rm
  eff} / h^{-1} M_\odot) = (13.48, 13.54, 13.61, 13.68)$.  For
$\sigma_8 = 0.9$, the results were $(13.72, 13.79, 13.85, 13.91)$.  In
all cases the minimum values of $\chi^2$ were similar to those
obtained for our default value of $\sigma_8$, suggesting that there is
a strong degeneracy between $\sigma_8$ and the halo model parameters
used in this analysis.  The changing $\sigma_8$ affected the
best-fitting value of $M_0$ much more strongly than the value of
$\beta$.  We also tried lowering the value of the scalar spectral
index of the primordial power spectrum from $n_{\rm s} = 1$ to $n_{\rm
  s} = 0.95$, as supported by recent observations of the Cosmic
Microwave Background radiation (Spergel et al.\ 2007).  Assuming
$\sigma_8 = 0.8$, the best-fitting effective masses are $(13.59,
13.66, 13.72, 13.78)$ which do not differ significantly from our
default values presented in Table \ref{tabhalofit}.

\section{Comparison to previous work}
\label{secprev}

Several previous studies have fitted halo model parameters to
populations of red galaxies, for example: Magliocchetti \& Porciani
(2003, 2dFGRS); Zehavi et al.\ (2004, 2005b, SDSS); Collister \& Lahav
(2005, 2dFGRS groups catalogue); Phleps et al.\ (2006, COMBO-17
survey); and White et al.\ (2007, NDWFS).  We make comparisons to
these analyses below, where possible.

Halo model fits to the 2dFGRS galaxy correlation function for
late-type and early-type galaxies were performed by Magliocchetti \&
Porciani (2003).  In addition, Collister \& Lahav (2005) directly
investigated the distribution of galaxies within 2dFGRS groups.  These
two studies produced a reasonably consistent measurement of the slope
$\beta \approx 1$ of the HOD at high masses for early-type galaxies.
Our best-fitting slope, $\beta = 1.5 \rightarrow 2.0$, is much higher
due to two factors: (1) the significantly higher luminosity of our
galaxy samples; (2) we make the distinction between central and
satellite galaxies, separating out a central galaxy contribution $N_c
\approx 1$ at high masses. This latter has the effect of significantly
steepening the slope of the power-law HOD fitted to the remaining
satellite galaxies (which in fact contribute only $5-10\%$ of our
sample, as noted in Section \ref{secparfit}).  In other words, we
effectively fit a model $N = N_c(1 + N_s) \approx 1 + (M/M_0)^\beta$
at high masses, rather than $N = (M/M_0)^\beta$.

Zehavi et al.\ (2004) analyzed a luminous subset of galaxies from the
SDSS ``main'' spectroscopic database with $M_r < -21$ and mean
redshift $z \approx 0.1$.  They found that a HOD of the form
\begin{eqnarray}
<N|M> &=& 0 \; \; \; (M < M_{\rm cut}) \nonumber \\ &=& 1 \; \; \;
(M_{\rm cut} < M < M_0) \nonumber \\ &=& (M/M_0)^\beta \; \; \; (M >
M_0)
\end{eqnarray}
produced a good fit to the clustering data, where $M_{\rm cut} =
10^{12.79} \, h^{-1} \, M_\odot$, $M_0 = 10^{13.68} \, h^{-1} \,
M_\odot$ and $\beta = 0.89$.  The effective mass corresponding to
these parameters is $M_{\rm eff} = 10^{13.83} \, h^{-1} M_\odot$ (for
their choice of $\sigma_8 = 0.9$).  The number density of the Zehavi
et al.\ (2004) sample is $n_g = 9.9 \times 10^{-4} \, h^3$ Mpc$^{-3}$,
which is a factor of $2-3$ higher than our study.  Although the
effective masses are similar, we note that the redshift difference
between the Zehavi et al.\ (2004) sample and ours may be important;
the number density of a sample of dark matter haloes of fixed mass
increases with decreasing redshift owing to the growth of structure.
The difference in the best-fitting value of the power-law slope
between Zehavi et al.\ and our analysis is again connected to the
different forms of HOD fitted (owing to our central galaxy
contribution, as discussed above).

Zehavi et al.\ (2005b) presented an extended analysis of the SDSS data
in which central and satellite galaxy contributions are considered
separately.  Their default model includes a sharp cut-off for the
central galaxy HOD at a fixed mass, rather than our ``softened''
transition from 0 to 1 galaxies.  They find that the slope of the
power-law satellite HOD increases systematically with luminosity in a
manner entirely consistent with our high-luminosity measurements of
$\beta = 1.5 \rightarrow 2.0$.  In addition, Zehavi et al.\ (2005b)
note that the step function for $<N_c|M>$ produces a poor fit to the
data in their highest-luminosity bin, consistent with our requirement
for a softened transition parameterized by $\sigma_{\rm cut}$.  They
also find, in agreement with our analysis, that the great majority of
luminous galaxies are central galaxies of their host dark matter
haloes, rather than satellites in more massive systems.  A low ($\la
10\%$) satellite fraction for the most luminous elliptical galaxies is
also found in galaxy-galaxy lensing studies (Seljak et al.\ 2005;
Mandelbaum et al.\ 2006a) and other clustering studies (Tinker et al.\
2007; van den Bosch et al.\ 2007).

Phleps et al.\ studied various populations of galaxies in the COMBO-17
survey at a mean redshift $\overline{z} = 0.6$ which is similar to our
dataset.  For red-sequence galaxies, Phleps et al.\ quote an effective
halo mass for their best-fitting model of $M_{\rm eff} = 10^{13.2} \,
h^{-1} M_\odot$, whereas we find $M_{\rm eff} = 10^{13.7} \, h^{-1}
M_\odot$ (Table \ref{tabhalofit}).  This apparently large discrepancy
is caused by the significant difference in the luminosity threshold of
the two samples: the number density of our LRG catalogue is more than
an order of magnitude smaller (there is also a difference in the
assumed value of $\sigma_8$).

White et al.\ (2007) fitted a Halo Occupation Distribution model to
the clustering of Luminous Red Galaxies in the NOAO Deep Wide-Field
Survey (NDWFS) Bootes field of 9 deg$^2$, analyzed in redshift slices
between $z = 0.4$ and $z = 1.0$.  The luminosity thresholds are fixed
such that the galaxy number density in each redshift slice is $10^{-3}
\, h^3$ Mpc$^{-3}$, exceeding our sample by a factor $\approx 3$ at $z
= 0.5$.  White et al.\ demonstrated that the clustering of the $z =
0.5$ sample cannot be accounted for by simple passive evolution of the
$z = 0.9$ sample, but rather there must be merging or disruption of
the most luminous satellite galaxies in massive haloes.  The
best-fitting satellite fraction in the NDWFS sample is found to be
$18\%$, a little higher than the results of our study, but consistent
with a trend in which satellite fraction decreases with increasing
luminosity.

In conclusion, our halo model parameter measurements appear broadly
consistent with previous work, allowing for differing luminosity
thresholds.  A fully consistent comparison of our analysis at $z
\approx 0.55$ with results at $z \approx 0$ is beyond the scope of
this work, owing to the differing forms of halo occupation
distribution assumed by different authors, but a topic worthy of
further investigation.

Measurement of the 3-point clustering functions will add further
insight into the LRG clustering properties.  Recent work by Kulkarni
et al.\ (2007), analyzing the SDSS spectroscopic LRG sample at $z
\approx 0.35$, favoured a shallower slope for the satellite HOD,
$\beta \approx 1.4$, with a higher satellite fraction of $17\%$.
Further study is required to understand these differences.

\section{Tests for systematic photometric errors}
\label{secsys}

We performed a series of tests for potential systematic photometric
errors that may affect our clustering results.  Following the
discussion in Blake et al.\ (2007), we compared the angular
correlation function measured for the ``default'' sample with that
obtained by restricting or extending the galaxy selection in the
following ways:
\begin{itemize}

\item Exclusion of areas of high dust extinction ($> 0.1$ mag).

\item Exclusion of areas of poor astronomical seeing ($> 1.5$ arcsec).

\item Exclusion of areas lying in the overlap regions between survey
  stripes.

\item Exclusion of areas in the vicinity of very bright objects
  (circular masks of radius 1 arcmin around objects with $i < 12$).

\item Variations in the star-galaxy separation criteria.  This is
  quantified by the coefficient $\delta_{\rm sg}$ in Blake et
  al.\ (2007) and Collister et al.\ (2007), which fixes the
  aggressiveness of the star-galaxy separation in the neural network.
  Our default choice is $\delta_{\rm sg} > 0.2$, which results in a
  level of stellar contamination of $1.5\%$ with the loss of only
  $\sim 0.1\%$ of the genuine galaxies (see Fig.\ 13 in Collister et
  al.\ 2007).  We also tried $\delta_{\rm sg} > 0$ (no additional
  star-galaxy separation in the neural network; stellar contamination
  $4.4\%$) and $\delta_{\rm sg} > 0.8$ (stellar contamination $0.4\%$;
  loss of $1.2\%$ genuine galaxies).

\end{itemize}
We refer the reader to Blake et al.\ (2007) for a more thorough
discussion of the possible effects of these systematic errors.

\begin{figure*}
\center
\epsfig{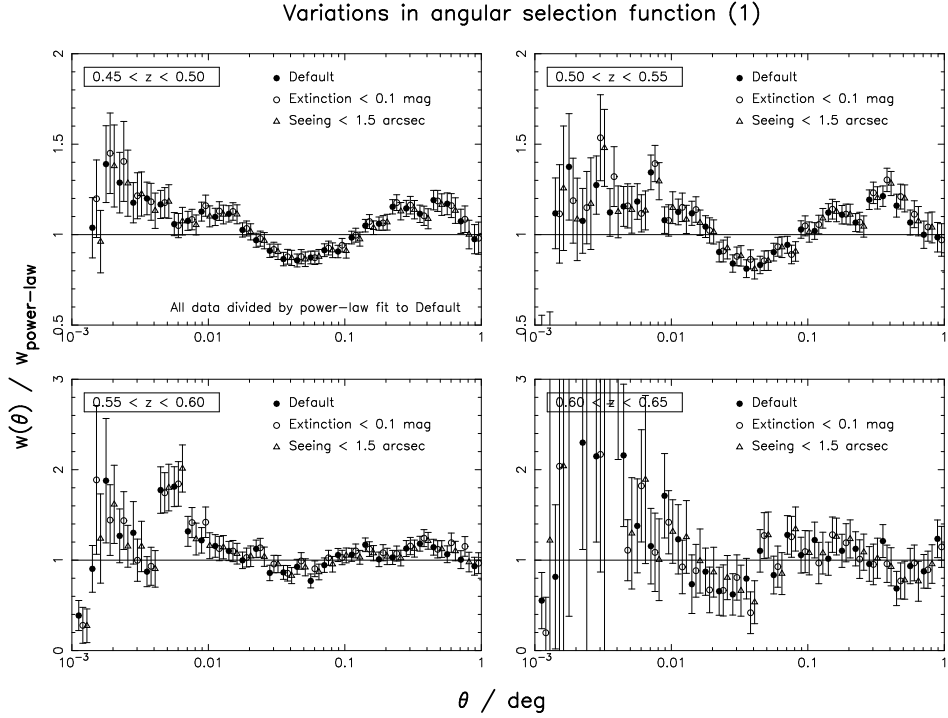}
\caption{The dependence of the angular correlation function
  measurement in 4 redshift slices on varying {\it dust extinction}
  and {\it astronomical seeing}.  Results from the default catalogue
  are compared to an analysis restricting the regions analyzed to (1)
  a maximum dust extinction of $0.1$ mag, or (2) a maximum seeing of
  $1.5$ arcsec.  The $y$-axis is the angular correlation function
  $w(\theta)$ divided by a power-law fit to the default model.}
\label{figsys1}
\end{figure*}

\begin{figure*}
\center
\epsfig{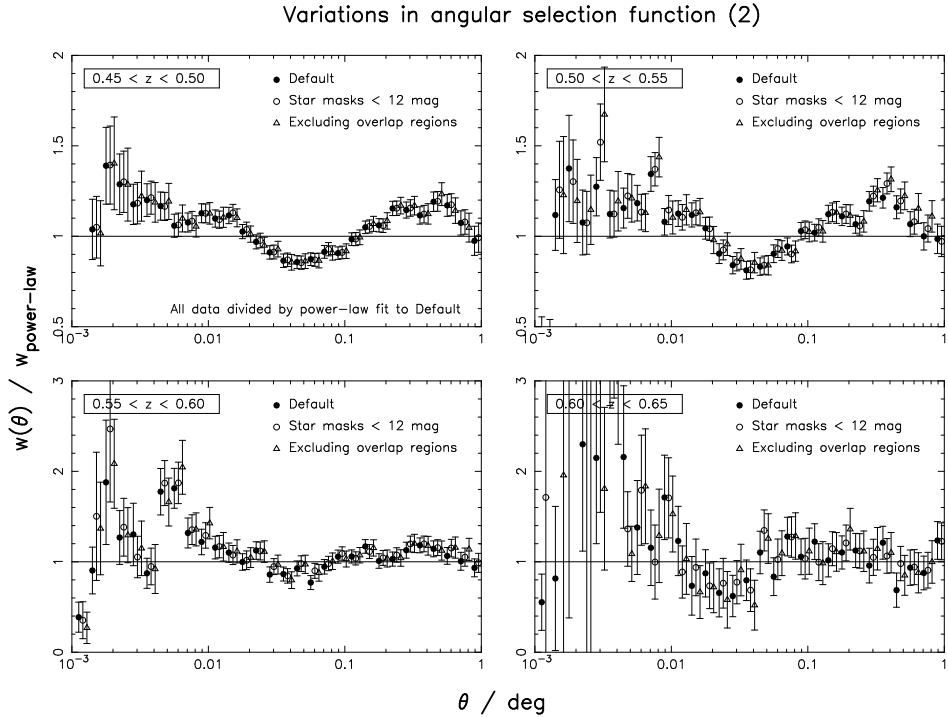}
\caption{The dependence of the angular correlation function
  measurement in 4 redshift slices on the presence of {\it bright
    objects} and {\it stripe overlap regions}.  Results from the
  default catalogue are compared to (1) an analysis placing circular
  masks of radius 1 arcmin around all objects with $i$-band magnitudes
  brighter than 12, and (2) an analysis excluding overlap regions
  between stripes.  The results are displayed in the same manner as
  Figure \ref{figsys1}.}
\label{figsys2}
\end{figure*}

\begin{figure*}
\center
\epsfig{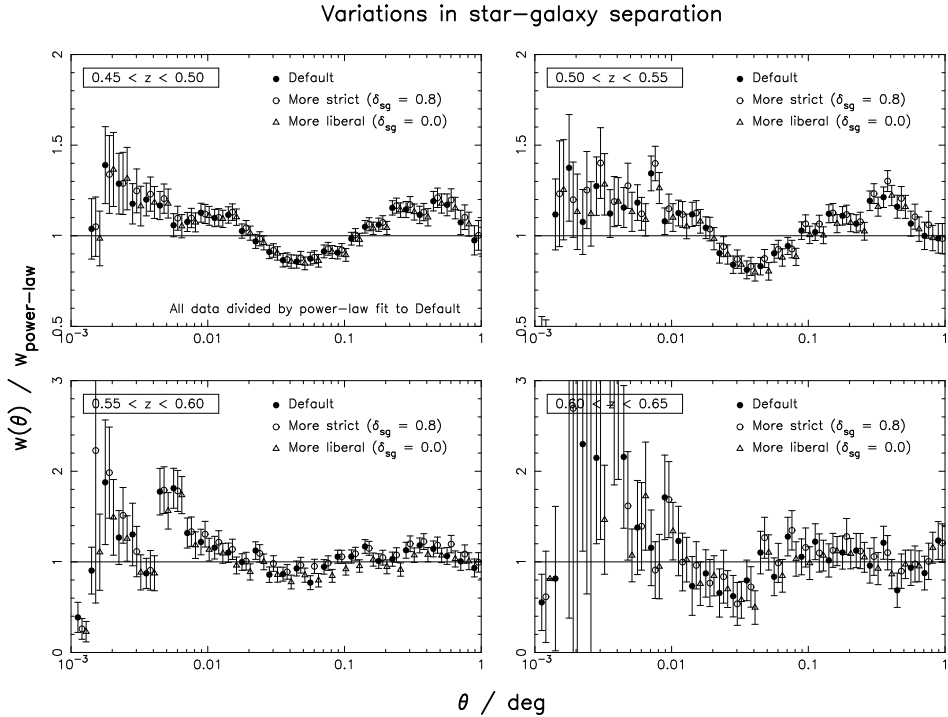}
\caption{The dependence of the angular correlation function
  measurement in 4 redshift slices on the star-galaxy separation
  criteria.  Results from the default catalogue are compared to
  analyses changing the value of the co-efficient $\delta_{\rm sg}$
  defined in Blake et al.\ (2007) and Collister et al.\ (2007), which
  controls the aggressiveness of the star-galaxy separation.  The
  results are displayed in the same manner as Figure \ref{figsys1}.}
\label{figsys3}
\end{figure*}

Our results are presented in Figures \ref{figsys1}, \ref{figsys2} and
\ref{figsys3}.  Each plot is composed of four panels, one for each
redshift slice.  In each panel we show the angular correlation
function for the default sample together with that corresponding to a
change in the galaxy selection criteria.  We divide all the
correlation functions by a power-law fit to the default measurement to
render the results more easily comparable.

We conclude from Figures \ref{figsys1} and \ref{figsys2} that our
results are robust against the details of the angular selection
function: varying dust extinction, seeing, overlap regions and bright
object masks all have little effect on the measured correlation
function.  Figure \ref{figsys3} reveals that the details of the
star-galaxy separation affect the amplitude of the measured
correlation function although not (to first order) the shape.  This
amplitude shift is already encoded in the stellar contamination factor
$(1-f)^2$.  In no case does a change in the galaxy selection alter the
detectability or shape of the halo model signature.

\section{Conclusions}
\label{secconc}

We have measured the angular correlation function of LRGs in the SDSS
imaging survey, using accurate photometric redshifts to divide the
galaxies into narrow redshift slices and create volume-limited
samples.  We find that:

\begin{itemize}

\item A canonical power-law fit provides a poor description of the
  small-scale angular correlation function, although the best-fitting
  slope $w(\theta) \propto \theta^{-0.95}$ agrees well with previous
  studies of Luminous Red Galaxies.

\item The halo model of galaxy clustering, composed of separate 1-halo
  and 2-halo contributions, produces a good fit to the deviations from
  a power-law.  We assume a halo occupation distribution with separate
  components for central and satellite galaxies, implementing
  realistic models for scale-dependent bias, halo exclusion and
  non-linear growth of structure.  We find that the HOD for central
  galaxies requires a ``soft'' transition from zero to one central
  galaxy, as opposed to a step function, to re-produce the
  observations.  The functional form $N_c = 0.5 \{ 1 + {\rm erf} [
    {\rm log}_{10} (M/M_{\rm cut}) / \sigma_{\rm cut} ] \}$ provides a
  good fit for central galaxies, combined with a power-law HOD for
  satellite galaxies $N_s = (M/M_0)^\beta$.  One parameter of the
  model ($M_{\rm cut}$) is fixed by the overall number density of the
  galaxy sample; hence this halo model contains 3 variable parameters
  ($\sigma_{\rm cut}$, $M_0$ and $\beta$), just one more than a simple
  power law.  Allowing the concentration parameter $c$ to vary, or
  including a more sophisticated HOD for satellite galaxies, does not
  improve the model fits.

\item The slope $\beta$ of the power-law HOD for satellite galaxies is
  a strong function of luminosity, increasing to $\beta \approx 2$ for
  our most luminous sample.  This is consistent with extrapolating the
  variation of $\beta$ with luminosity in local samples (Zehavi et
  al.\ 2005b).  The physical implication of this result is that haloes
  of higher mass have greater relative efficiency at producing
  high-luminosity satellites.  We find no variation of $\beta$ with
  redshift across our sample (from $z = 0.45$ to $z = 0.65$).

\item The best-fitting width $\sigma_{\rm cut}$ of the transition from
  zero to one central galaxy with increasing mass is in the range
  $\sigma_{\rm cut} = 0.1 \rightarrow 0.5$ (this is the parameter
  least-well constrained by our analysis).  The physical implication
  of this result is a scatter between central galaxy luminosity and
  host halo mass.  This scatter results from galaxy formation physics
  and from the photo-$z$ error in the conversion of apparent magnitude
  to luminosity.

\item The halo model fits describe how Luminous Red Galaxies populate
  dark matter haloes as a function of their mass $M$.  The average
  number of our galaxy sample hosted by a halo of mass $M = 10^{14.5}
  \, h^{-1} \, M_\odot$ is $(5.5, 4.1, 2.6, 1.4)$ in the four redshift
  slices, decreasing systematically with redshift as the threshold
  luminosity increases.  Broadly speaking, the effect of increasing
  luminosity is to shift the HOD uniformly to higher masses without
  significantly changing its shape.  The large majority of galaxies in
  our sample are classified as central galaxies of their host dark
  matter haloes, rather than satellites in more massive systems, in
  agreement with previous studies of galaxy-galaxy lensing and
  clustering of the most luminous galaxies.  The satellite fraction
  varies in the range $3\% \rightarrow 12\%$ across the redshift
  slices.

\item The halo model fits provide robust predictions of the average
  linear bias of the LRGs on large scales and the effective mass of
  their host dark matter haloes.  The resulting amplitude of the
  linear bias ($b_g = 1.9 \rightarrow 2.6$, increasing with redshift,
  assuming a normalization of the matter power spectrum $\sigma_8 =
  0.8$) agrees well with fits to the large-scale power spectrum (Blake
  et al.\ 2007).  The effective halo mass ($M_{\rm eff} = 10^{13.6
    \rightarrow 13.8} \, h^{-1} M_\odot$) provides a quantitative
  statement of how the LRGs trace the underlying dark matter haloes.
  The value of $M_{\rm eff}$ has a significant dependence on
  $\sigma_8$, and a weak dependence on the slope of the primordial
  scalar index $n_{\rm s}$: the effective mass increases by $\Delta
  {\rm log}_{10}(M_{\rm eff}/h^{-1} M_\odot) \approx 0.1$ when the
  value of $\sigma_8$ is increased from $0.8$ to $0.9$, and decreases
  by $\Delta {\rm log}_{10}(M_{\rm eff}/h^{-1} M_\odot) \approx 0.02$
  when $n_{\rm s}$ is decreased from $1.0$ to $0.95$.

\end{itemize}

Future studies will explore joint fits of the cosmological parameters
and halo model parameters (Abazajian et al.\ 2005; Zheng \& Weinberg
2007); direct measurement of halo occupation via a cluster and
group-finding analysis of the photo-$z$ catalogue; a consistent
comparison of clustering of Luminous Red Galaxies at $z \approx 0.5$
and at $z \approx 0$; and testing the halo model further via 3-point
clustering statistics and higher moments.

\section*{Acknowledgments}

We are very grateful to an anonymous referee for extremely detailed
comments that greatly improved this paper.  We also thank Zheng Zheng
for providing very helpful advice and code comparisons for the halo
occupation distribution modelling.  We thank Sarah Bridle, Filipe
Abdalla, Ravi Sheth and Michael Brown for useful discussions.  CB
acknowledges support from the Izaak Walton Killam Memorial Fund for
Advanced Studies and from the Canadian Institute for Theoretical
Astrophysics National Fellowship programme.  OL acknowledges a PPARC
Senior Research Fellowship.

Funding for the Sloan Digital Sky Survey (SDSS) has been provided by
the Alfred P. Sloan Foundation, the Participating Institutions, the
National Aeronautics and Space Administration, the National Science
Foundation, the U.S. Department of Energy, the Japanese
Monbukagakusho, and the Max Planck Society. The SDSS Web site is
{\tt http://www.sdss.org/}.

The SDSS is managed by the Astrophysical Research Consortium (ARC) for
the Participating Institutions. The Participating Institutions are The
University of Chicago, Fermilab, the Institute for Advanced Study, the
Japan Participation Group, The Johns Hopkins University, the Korean
Scientist Group, Los Alamos National Laboratory, the
Max-Planck-Institute for Astronomy (MPIA), the Max-Planck-Institute
for Astrophysics (MPA), New Mexico State University, University of
Pittsburgh, University of Portsmouth, Princeton University, the United
States Naval Observatory, and the University of Washington.


\begin{thebibliography}{}
\bibitem{1} Abazajian K. et al., 2005, ApJ, 625, 613
\bibitem{2} Amendola L., Quercellini C., Giallongo E., 2005, MNRAS,
  357, 429
\bibitem{3} Benson A.J., Cole S., Frenk C.S., Baugh C.M., Lacey C.G.,
  2000, MNRAS, 311, 793
\bibitem{4} Berlind A.A., Weinberg D.H., 2002, ApJ, 575, 587
\bibitem{5} Berlind A.A. et al., 2003, ApJ, 593, 1
\bibitem{6} Blake C.A., Bridle S.L., 2005, MNRAS, 363, 1329
\bibitem{7} Blake C.A., Collister A., Bridle S.L., Lahav O., 2007,
  MNRAS, 374, 1527
\bibitem{8} Brown M.J.I., Dey A., Jannuzi B.T., Lauer T.R., Tiede
  G.P., Mikles V.J., 2003, ApJ, 597, 225
\bibitem{9} Budavari T. et al., 2003, ApJ, 595, 59
\bibitem{10} Bullock J.S., Kolatt T.S., Sigad Y., Somerville R.S.,
  Kravtsov A.V., Klypin A.A., Primack J.R., Dekel A., 2001, MNRAS,
  321, 559
\bibitem{11} Cannon R. et al., 2006, MNRAS, 372, 425
\bibitem{12} Collister A., Lahav O., 2004, PASP, 116, 345
\bibitem{13} Collister A., Lahav O., 2005, MNRAS, 361, 415
\bibitem{14} Collister A. et al., 2007, MNRAS, 375, 68
\bibitem{15} Cooray A., Sheth R., 2002, Phys.Rep., 372, 1
\bibitem{16} Dolney D., Jain B., Takada M., 2004, MNRAS, 352, 1019
\bibitem{17} Eisenstein D.J., Hu W., 1998, ApJ, 518, 2
\bibitem{18} Eisenstein D.J. et al., 2001, AJ, 122, 2267
\bibitem{19} Eisenstein D.J. et al., 2005a, ApJ, 619, 178
\bibitem{20} Eisenstein D.J. et al., 2005b, ApJ, 633, 560
\bibitem{21} Firth A.E., Lahav O., Somerville R.S., 2003, MNRAS, 339,
  1195
\bibitem{22} Fry J.N., 1996, ApJ, 461, 65
\bibitem{23} Jenkins A., Frenk C.S., White S.D.M., Colberg J.M., Cole
  S., Evrard A.E., Couchman H.M.P., Yoshida N., 2001, MNRAS, 321, 372
\bibitem{24} Hawkins E. et al., 2003, MNRAS, 346, 78
\bibitem{25} Hogg D.W. et al., 2003, ApJ, 585, L5
\bibitem{26} Kauffmann G., Nusser A., Steinmetz M., 1997, MNRAS, 286, 795
\bibitem{27} Kravtsov A.V., Berlind A.A., Wechsler R.H., Klypin A.A.,
  Gottloeber S., Allgood B., Primack J.R., 2004, ApJ, 609, 35
\bibitem{28} Kulkarni G.V., Nichol R.C., Sheth R.K., Seo H.-J.,
  Eisenstein D.J., Gray A., 2007, MNRAS, 378, 1196
\bibitem{29} Lahav O. et al., 2002, MNRAS, 333, 961
\bibitem{30} Landy S.D., Szalay A.S., 1993, ApJ, 412, 64
\bibitem{31} Lewis A., Challinor A., Lasenby A., 2000, ApJ, 538, 473
\bibitem{32} Madgwick D.S. et al., 2003, MNRAS, 344, 847
\bibitem{33} Magliocchetti M., Porciani C., 2003, MNRAS, 346, 186
\bibitem{34} Mandelbaum R., Seljak U., Kauffmann G., Hirata C.M.,
  Brinkmann J., 2006a, MNRAS, 368, 715
\bibitem{35} Mandelbaum R., Seljak U., Cool R.J., Blanton M., Hirata
  C.M., Brinkmann J., 2006b, MNRAS, 372, 758
\bibitem{36} Navarro J.F., Frenk C.S., White S.D.M., 1997, ApJ, 490, 493
\bibitem{37} Norberg P. et al., 2002, MNRAS, 332, 827
\bibitem{38} Padmanabhan N. et al., 2005, MNRAS, 359, 237
\bibitem{39} Padmanabhan N. et al., 2007, MNRAS, 378, 852
\bibitem{40} Peacock J.A., Smith R.E., 2000, MNRAS, 318, 1144
\bibitem{41} Peebles P.J.E., 1980, {\it The Large-Scale Structure of
  the Universe,} Princeton University Press, Princeton
\bibitem{42} Phleps S., Peacock J.A., Meisenheimer K., Wolf C., 2006,
  A\&A, 457, 145
\bibitem{43} Press W.H., Schechter P., 1974, ApJ, 187, 425
\bibitem{44} Scoccimarro R., Sheth R.K., Hui L., Jain B., 2001, ApJ,
  546, 20
\bibitem{45} Seljak U., 2000, MNRAS, 318, 203
\bibitem{46} Seljak U. et al., 2005, Phys. Rev. D, 71, 43511
\bibitem{47} Seo H., Eisenstein D.J., 2003, ApJ, 598, 720
\bibitem{48} Sheth R.K., Tormen G., 1999, MNRAS, 308, 119
\bibitem{49} Sheth R.K., Mo H.J., Tormen G., 2001, MNRAS, 323, 1
\bibitem{50} Smith R.E. et al., 2003, MNRAS, 341, 1311
\bibitem{51} Spergel D.N. et al., 2007, ApJS, 170, 377
\bibitem{52} Tinker J.L., Weinberg D.H., Zheng Z., Zehavi I., 2005,
  ApJ, 631, 41
\bibitem{53} Tinker J.L., Norberg P., Weinberg D.H., Warren M.S.,
  2007, ApJ, 659, 877
\bibitem{54} van den Bosch F.C., Yang X., Mo H.J., Weinmann S.M.,
  Maccio A., More S., Cacciato M., Skibba R., Xi K., 2007, MNRAS, 376,
  841
\bibitem{55} Wake D.A. et al., 2006, MNRAS, 372, 537
\bibitem{56} White M., Zheng Z., Brown M.J.I., Dey A., Jannuzi B.T.,
  2007, ApJL, 655, 69
\bibitem{57} York D.G. et al., 2000, AJ, 120, 1579
\bibitem{58} Zehavi I. et al., 2004, ApJ, 608, 16
\bibitem{59} Zehavi I. et al., 2005a, ApJ, 621, 22
\bibitem{60} Zehavi I. et al., 2005b, ApJ, 630, 1
\bibitem{61} Zhan H., Knox L., Tyson J.A., Margoniner V., 2006, ApJ,
  640, 8
\bibitem{62} Zheng Z., 2004, ApJ, 610, 61
\bibitem{63} Zheng Z. et al., 2005, ApJ, 633, 791
\bibitem{64} Zheng Z., Weinberg D.H., 2007, ApJ, 659, 1
\end{thebibliography}
\end{document}